\documentclass[aps,prd,showpacs,floatfix,preprint]{revtex4}
\usepackage{amsmath,bm}
\usepackage{graphicx}
\usepackage{epsfig}
\begin{document}

\title{Probing dense matter in neutron stars with axial w-modes}

\author{ Debarati Chatterjee and Debades Bandyopadhyay}
\affiliation{Theory Division and Centre for Astroparticle Physics,
Saha Institute of Nuclear Physics, 1/AF Bidhannagar, 
Kolkata-700064, India}

\begin{abstract}
We study the problem of extracting information about composition and equation
of state of dense matter in neutron star interior using axial w-modes. We 
determine complex frequencies of axial w-modes for a set of equations of state
involving hyperons as well as Bose-Einstein condensates of antikaons adopting 
the continued fraction method. Hyperons and antikaon condensates result in 
softer equations of state leading to higher frequencies of first axial w-modes 
than that of nuclear matter case, whereas the opposite happens in case of 
damping times. The presence of condensates may lead to the appearance of a new 
stable branch of superdense stars beyond the neutron star branch called the 
third family. The existence of same mass compact stars in both branches are 
known as neutron star twins. Further investigation of twins reveal that first 
axial w-mode frequencies of superdense stars in the third family are higher 
than those of the corresponding twins in the neutron star branch. 
\pacs{97.60.Jd, 26.60.-c, 04.40.Dg}
\end{abstract}

\maketitle

\section{Introduction}

Shortly after the discovery of pulsars, the study of dense matter in 
the core of neutron stars gained momentum \cite{Gle}. With the advent of 
satellite based observatories such as Einstein, ROSAT, Hubble space telescope 
XMM-Newton, and Chandra x-ray observatory, the study of neutron stars has 
entered into a
new era. Observations using these facilities as well as other facilities are 
pouring in very exciting data on neutron stars. It is now possible to 
estimate masses, radii, moment of inertia and surface temperatures of neutron 
stars from the observations. Those findings ,in turn, may shed light on the 
gross properties of cold dense matter far off from normal nuclear matter 
density in neutron star interior. Such a cold dense matter in the core of a 
neutron star, so far, has not been produced in laboratories on the earth. 
Therefore neutron stars are very useful laboratories for the study of highly  
dense matter in its core.   

On the other hand, there is a growing interplay between the physics of dense 
matter in relativistic heavy ion collisions and that of neutron stars.
Though the Quantum Chromodynamics predicts a very rich phase structure of dense
matter, we can only probe a small region of it in the laboratories. 
The study of dense matter in heavy ion collisions reveals many new 
and interesting results such as the modifications of hadron 
properties in dense medium, the properties of strange matter including 
hyperons and (anti)kaons, the strength of attractive antikaon-nucleon 
interaction and the formation of quark-gluon plasma. These 
empirical informations from heavy 
ion collisions may be useful in understanding dense matter in neutron star 
interior. Heavy ion experiments in the upcoming Facility for Antiproton and 
Ion Research (FAIR) at GSI, Germany would produce dense matter with baryon 
density a few times normal nuclear matter density and temperature a few tens 
of MeV and provide us important information about dense matter. 
  
Gravitational wave astrophysics opens a new window to probe the neutron star 
interior \cite{Owe}. Neutron stars are accompanied by strong gravitational 
field. Non-radial oscillations of neutron stars generate gravitational waves. 
Such a situation may arise in neutron star glitch, merger of compact stars,
a phase transition in the core and gravitational collapse of a massive star.  
Some large interferometric gravitational wave detectors such as
LIGO in USA, VIRGO in Italy and GEO600 in Germany are now operational. A 
neutron star has a large number of families of pulsation modes 
\cite{And,And01}. Those modes are
classified according to restoring forces acting on a perturbed fluid element.
Important modes among them are fundamental f-mode associated with global
oscillations of the fluid, g-mode due to buoyancy and p-mode due to pressure
gradient. General relativity also predicts w-modes which are space-time modes.
Unlike other pulsation modes of neutron stars,
w-modes do not have counterparts in Newtonian gravity. Equations governing
perturbation of a static and non-rotating neutron star are decomposed into two 
sets depending on how perturbations transform under parity. Polar perturbation
corresponds to even parity $(-1)^l$ and axial perturbation relates to odd 
parity $(-1)^{l+1}$ . Various groups \cite{Chan1,Chan2,Kok,Lei,Val} 
investigated w-modes in great details and found interesting varieties of axial 
w-modes \cite{Kok,Lei}. It is to be noted that w-modes have higher frequencies 
than those of $f$, $g$ and $p$ modes.

In general relativity,  oscillations of neutron stars are damped by 
gravitational
radiation. Consequently, the corresponding oscillatory modes are quasi-normal
modes and their frequencies are complex.
Once gravitational waves from a neutron star are detected, the spectral analysis
of the detected signal would tell us the frequency and damping time of the 
quasi-normal mode responsible for gravitational waves. Now the question is how 
we determine the mass, radius and properties of dense matter from the frequency
and damping time of the mode. This is achieved by solving the "inverse problem" 
\cite{And,Ben1,Ben2,Tsu}. To solve this problem, we
have to compute frequencies and damping times of modes that could generate
strong gravitational waves using different compositions and equations of state 
and compare them with those obtained from observations. This way we could probe
the neutron star interior.  

Various novel phases of matter might appear in neutron star interior because of
large baryon and lepton chemical potentials there. Glendenning \cite{Gle}
predicted the presence of hyperons in neutron stars. Kaplan and Nelson 
discussed Bose-Einstein condensation of $K^-$ mesons in neutron star matter 
due to strongly attractive antikaon-nucleon interaction \cite{Kap}. The 
formation of quark matter is another possibility in neutron star core 
\cite{Far}.   
In this paper, we are motivated to investigate the role of hyperons and 
antikaon condensation on frequencies of quasi-normal modes of neutron stars.  
In particular, we would investigate how the strength of 
antikaon optical potential in nuclear matter which is crucial for antikaon
condensation in neutron stars, affects axial w-modes.
In this connection, we want to calculate the frequency 
and damping time of axial w-mode of a neutron star including Bose-Einstein 
condensate of negatively charged kaons for a set of values of antikaon optical
potential depth. Further we investigate the role of hyperons on frequency and
damping time of axial w-modes. 

This paper is organised in the following way. The formalism is discussed in
section II. Results are explained in section III. Section IV provides summary 
and conclusions.
 
\section{Formalism}
Chandrasekhar and Ferrari showed how the problem non-radial oscillations of a 
compact star could be reduced to the scattering of incident gravitational
wave by the static space-time of spherically symmetric star \cite{Chan1,Chan2}.
Here we are interested in axial w-modes of neutron stars. The axial 
perturbation of the compact star is described by the metric 
\cite{Chan1,Chan2,Val}
\begin{equation}
ds^2 = e^{2\nu} {(dt)}^2 - e^{2\psi} {(d\phi - q_2 dr - q_3 d\theta 
- \omega dt)}^2 - e^{2\mu_2} {(dr)}^2 - e^{2\mu_3} {(d\theta)}^2,
\end{equation}  
where metric functions $\nu$, $\psi$, $\mu_2$ and $\mu_3$ retain their 
unperturbed values and $\omega$, $q_2$ and $q_3$ give rise to the 
perturbations. 
The radial evolution of equations for axial perturbations is described by the
following by the Schr\"odinger like equation with a potential barrier $V(r)$
\begin{eqnarray}
\left(\frac{d^2}{dr_*^2} + {\omega}^2 - V(r) \right) Z_l = 0~,
\label{ZR}
\\
V(r) = \frac{e^{2\nu}}{r^3} \left(l(l+1)r + r^3 [\epsilon(r) - P(r)]
-6 M(r)\right)~,\\
\frac{d\nu}{dr} = -{\frac{1}{\left(\epsilon(r) + P(r) \right)}}
{\frac{dP}{dr}}~, 
\end{eqnarray}
where the tortoise coordinate is $r_* = \int_0^r e^{-\nu + \mu_2} dr$, 
$M(r)$ is the mass enclosed in a radius $r$, $\omega$ is the complex frequency 
of the mode and $\epsilon$ and $P$ are energy density and pressure 
respectively. Both energy and pressure outside the star become zero and the
Eq. (\ref{ZR}) reduces to  the Regge-Wheeler equation with
\begin{equation}
V(r) = \left(1 - \frac{2M}{r} \right) \left(\frac{l(l+1)}{r^2} - \frac{6M}{r^3}
\right)~,
\end{equation}
and $r^* = r + 2M ln(r-2M)$.
We determine frequencies of quasi-normal modes by solving Eq. (\ref{ZR}) 
with appropriate $V(r)$ inside and outside the neutron star and 
boundary conditions such that the solution is regular at the center, continuous
at the surface and behaves as a purely outgoing wave at infinity. 

Now we focus on the construction of equation of state (EoS) (pressure 
versus energy density) within relativistic field theoretical models. In this 
paper, we consider EoS with different compositions and values of 
incompressibility of nuclear matter. Firstly we discuss the EoS including 
neutrons ($n$), protons ($p$), 
hyperons (H=$\Lambda$,$\Sigma^+$,$\Sigma^-$,$\Sigma^0$,$\Xi^-$,$\Xi^0$) and 
electrons ($e$) and muons($\mu$). 
The baryon-baryon interaction is mediated by the exchange of $\sigma$,
$\omega$, $\rho$ mesons and for hyperon-hyperon interaction, two
additional mesons - scalar meson $f_0$(975) (denoted hereafter as 
$\sigma^*$) and vector meson $\phi$(1020) \cite{Mis} are incorporated.
Therefore the Lagrangian density for baryons in the charge neutral and 
$\beta$-equilibrated hadronic phase is given by
\begin{eqnarray}
{\cal L}_B &=& \sum_B \bar\psi_{B}\left(i\gamma_\mu{\partial^\mu} - m_B
+ g_{\sigma B} \sigma - g_{\omega B} \gamma_\mu \omega^\mu
- g_{\rho B}
\gamma_\mu{\mbox{\boldmath t}}_B \cdot
{\mbox{\boldmath $\rho$}}^\mu \right)\psi_B\nonumber\\
&& + \frac{1}{2}\left( \partial_\mu \sigma\partial^\mu \sigma
- m_\sigma^2 \sigma^2\right) - U(\sigma) \nonumber\\
&& -\frac{1}{4} \omega_{\mu\nu}\omega^{\mu\nu}
+\frac{1}{2}m_\omega^2 \omega_\mu \omega^\mu
- \frac{1}{4}{\mbox {\boldmath $\rho$}}_{\mu\nu} \cdot
{\mbox {\boldmath $\rho$}}^{\mu\nu}
+ \frac{1}{2}m_\rho^2 {\mbox {\boldmath $\rho$}}_\mu \cdot
{\mbox {\boldmath $\rho$}}^\mu + {\cal L}_{YY}~.
\label{BB}
\end{eqnarray}
The isospin multiplets for baryons B $=$ N, $\Lambda$, $\Sigma$ and $\Xi$ are
represented by the Dirac spinor $\Psi_B$ with vacuum baryon mass $m_B$
and isospin operator ${\mbox {\boldmath t}}_B$ and $\omega_{\mu\nu}$ and
$\rho_{\mu\nu}$ are field strength tensors. The scalar
self-interaction term \cite{Bog} is
\begin{equation}
U(\sigma) = \frac{1}{3} g_2 \sigma^3 + \frac{1}{4} g_3 \sigma^4 ~.
\end{equation}
The Lagrangian density for hyperon-hyperon interaction (${\cal L}_{YY}$)
is given by
\begin{eqnarray}
{\cal L}_{YY} &=& \sum_B \bar\Psi_{B}\left(
g_{\sigma^* B} \sigma^* - g_{\phi B} \gamma_\mu \phi^\mu
\right)\Psi_B\nonumber\\
&& + \frac{1}{2}\left( \partial_\mu \sigma^*\partial^\mu \sigma^*
- m_{\sigma^*}^2 \sigma^{*2}\right)
-\frac{1}{4} \phi_{\mu\nu}\phi^{\mu\nu}
+\frac{1}{2}m_\phi^2 \phi_\mu \phi^\mu~.
\label{Lag}
\end{eqnarray}
As nucleons do not couple with strange mesons, 
$g_{\sigma^* N} = g_{\phi N} = 0$.

Next we discuss equations of state undergoing first order phase transitions 
from nuclear to $K^-$ condensed matter denoted by np$K^-$ and hyperon matter to
$\bar K$ ($K^-$, $\bar K^0$) condensed matter denoted by npH$\bar K$. 
Here we have pure hadronic phase described by Eq. (\ref{BB}) 
and antikaon condensed phases and the mixed phase of two pure phases. 
The constituents of $\beta$-equilibrated matter in both phases are neutrons, 
protons, hyperons, electrons and muons. Baryons are embedded in the condensate 
in the condensed phase. We describe the interaction of (anti)kaons in a 
relativistic field theoretical model.
The Lagrangian density for (anti)kaons in the minimal coupling scheme is
\cite{Gle98,Gle99,Pal,Bani1,Bani2}, 
\begin{equation}
{\cal L}_K = D^*_\mu{\bar K} D^\mu K - m_K^{* 2} {\bar K} K ~,
\end{equation}
where the covariant derivative is
$D_\mu = \partial_\mu + ig_{\omega K}{\omega_\mu} + ig_{\phi K}{\phi_\mu}
+ i g_{\rho K}
{\mbox{\boldmath t}}_K \cdot {\mbox{\boldmath $\rho$}}_\mu$ and
the effective mass of (anti)kaons is 
$m_K^* = m_K - g_{\sigma K} \sigma - g_{\sigma^* K} \sigma^*$.
The mixed phase of hadronic and $K^-$ condensed matter is governed by  the 
Gibbs conditions for thermodynamic equilibrium along with global charge and
baryon number conservation laws \cite{Gle92}. 
Working in the mean field approximation, we obtain energy density and pressure
in each pure phase and those are given by Ref.\cite{Bani2}. 

\section{Results and Discussion}
Nucleon-meson coupling constants obtained by reproducing saturation properties
of nuclear matter for two values of incompressibility $K=240$ and $K=300$ MeV,
are taken from Ref. \cite{Gle91}. These are known as GM sets. 
Hyperon-vector meson coupling constants are determined from SU(6) symmetry of
the quark model \cite{Mis,Dov,Sch94}. The scalar meson $\sigma$ meson
coupling to hyperons is calculated from the potential depth of a hyperon (Y)
\begin{equation}
U_Y^N (n_0) = - g_{\sigma Y} \sigma + g_{\omega Y} \omega_0~,
\end{equation}
in normal nuclear matter. Potential depths of hyperons in normal nuclear matter
are obtained from the analysis of hypernuclei data \cite{Dov,Chr,Fuk,Kha}. 
Similarly
hyperon-$\sigma^*$ coupling constants are estimated by fitting them to a
potential depth, ${U_{Y}^{(Y^{'})}}{(n_0)}$, for a hyperon (Y) in hyperon
matter. We adopt the values of hyperon potential depths for this calculation
as quoted in Ref. \cite{Bani2}. Next we determine 
kaon-vector meson coupling constants using the quark model and isospin counting
rule \cite{Gle98}. The scalar coupling with kaons is obtained from the real 
part of $K^-$ optical potential depth at normal nuclear matter density
\cite{Fri94,Fri99}
\begin{equation}
U_{\bar K} \left(n_0\right) = - g_{\sigma K}\sigma - g_{\omega K}\omega_0 ~,
\end{equation}
where $\sigma$ and $\omega_0$ are mean values of the meson fields.
We perform our calculation for antikaon condensation with $K=300$ MeV and two 
values of antikaon optical potential depth $U_{\bar K} (n_0) = -120, -160$ 
MeV.
Values of $g_{\sigma K}$ are recorded in Ref. \cite{Bani2}. Coupling constants 
$g_{\sigma^*K}$ and $g_{\phi K}$ are obtained from 
Refs.\cite{Mis,Bani2}. 

Now we present our results for different compositions and equations of state.
Gravitational mass for static neutron star as well as superdense star sequences
are shown with radius in Figure 1 for different values of incompressibility.
For $n$$p$ matter we find the EoS with $K=300$ MeV is stiffer than that of the 
case with $K=240$ MeV. Consequently the maximum mass is larger in the former
case. With the appearance of hyperons and/or antikaon condensates, equations of
state become softer. For $n$$p$$\Lambda$ matter and $K=300$ MeV, the threshold
density of $\Lambda$ hyperons is 2.3$n_0$. On the other hand, for $n$$p$$K^-$
matter with $K=300$ MeV, $K^-$ condensation sets in at 2.23$n_0$ and the 
mixed phase terminates at 3.59$n_0$ for $U_{\bar K}{(n_0)} =$ -160 whereas 
the phase transition for $U_{\bar K}(n_0) =$ -120 MeV begins at 3.05$n_0$ and
there is no mixed phase in this case. Stronger the attactive antikaon optical 
potential depth, softer is the EoS. Softer EoS involving hyperons and antikaon 
condensates lead to reduction in maximum masses than that of the $n$$p$ matter
as evident from Fig. 1. For $n$$p$H$K^-$$\bar K^0$
case with $K=300$ and $U_{\bar K}(n_0) = -160$ MeV, $\Lambda$ hyperons appear
at 2.51$n_0$ after the onset of $K^-$ condensation at 2.23$n_0$. 
The mixed phase ends at 4.0$n_0$ in this case. 
Further $\bar K^0$ condensation which is treated here as a second order phase 
transition, begins at 4.06$n_0$ \cite{Bani2}. Heavier strange baryons are 
populated
beyond 6$n_0$ \cite{Bani2}. For $n$$p$H$K^-$$\bar K^0$ case, it was noted that
after the stable neutron star branch, there was an unstable region followed by
another stable branch of compact stars \cite{Bani2}. As white dwarfs and 
neutron stars form first two families of compact stars, this branch of 
superdense stars beyond the neutron star branch is called the third family of
compact stars \cite{Ger}. Further the existence of non-identical stars having
same mass could be possible because of partial overlapping of the neutron star 
and third family branches. These pairs of compact stars are known as "neutron 
star twins" \cite{Ket,Scher}. Superdense stars in the third family branch have 
different compositions and smaller radii than their counterparts in the neutron
star branch \cite{Bani2}.           

Next we calculate the frequency ($\nu$) and damping time ($\tau$) from Eq. 
(\ref{ZR}) using equations of state described above. We exploit both the
continued fraction method \cite{Ben1} and integration method \cite{Kok} to find
complex eigenfrequency of the Schr\"odinger like equation (\ref{ZR}). Here we
present our results obtained by using the continued fraction method. The
frequency and damping time are related to the real and imaginary parts of
eigenfrequency through $\nu = \frac{32.26}{n} (M\omega_0)$ kHz and 
$\tau = 4.937 \times \frac{n}{M\omega_i}$ $\mu$s, where n=M/$M_{\odot}$ and 
$\omega_0$, $\omega_i$, M and M$_{\odot}$ are measured in units of km. 
Figures 2 and 3 display frequency and damping time of first axial w-mode as a 
function of neutron star compactness (M/R) respectively. In both figures, 
the dashed line corresponds to $n$$p$ matter with $K=240$ MeV whereas the solid 
line represents $n$$p$ case with $K=300$ MeV. We note that the softer EoS
leads to higher frequency of first axial w-mode for each M/R value as shown in 
Fig. 2 whereas it is the reverse for damping time as evident from Fig. 3. 

In the next paragraphs we discuss the role of exotic matter on frequency and
damping time of axial w-mode. Firstly we study axial w-modes of neutron stars
involving $n$$p$$\Lambda$ matter along with those of $n$$p$ matter for 
$K=300$ MeV.
Frequencies and damping times of first axial w-mode for both cases are exhibited
in Figures 4 and 5. The appearance of $\Lambda$ hyperons makes the EoS softer 
compared with the $n$$p$ case. This leads to higher frequencies of first axial 
modes for oscillating neutron stars including $\Lambda$ hyperons as shown by
the upper curve in Fig. 4 
whereas damping times in this case are given by the lower curve in Fig. 5.
Frequencies and damping times of first three $l=2$ axial w-modes corresponding
to the EoS involving $n$$p$$\Lambda$ matter with 
increasing compactness are recorded in Table I. 

We continue our investigation of first axial w-modes of neutron stars including
a first order $K^-$ condensate. These results are shown in Figures 6 and 7. We
have considered two values of antikaon optical potential depth at normal 
nuclear matter density $U_{\bar K} =$ -120, -160 MeV. For $U_{\bar K} =$ -160 
MeV, frequencies of first axial w-modes for neutron stars having a $K^-$ 
condensate are shown by the top curve whereas those of $n$$p$$K^-$ EoS for 
$U_{\bar K}(n_0) =$ -120 MeV have lower frequencies. This may be attributed to 
the fact that the EoS with $U_{\bar K} (n_0) =$ -160 is softer than that of
$U_{\bar K} (n_0) =$ -120 MeV. Similarly we find
that damping times for $U_{\bar K}(n_0) =$ -160 MeV are reduced than those of
$U_{\bar K}(n_0) =$ -120 MeV and $n$$p$ case.  
Frequencies and damping times of first three $l=2$ axial w-modes  of 
oscillating neutron stars including a $K^-$ condensate with 
increasing compactness are recorded in Table II. 

Frequencies and damping times of first axial w-modes of neutron star as well
as third 
family sequences as a function of compactness are shown in Figures 8 and 9.
In Fig. 8 frequencies of first axial w-modes corresponding to the neutron star
and third family branches are denoted by the bottom and top curves 
respectively. Further twins of superdense stars in the third family branch (top
curve) are represented by filled squares in the neutron star branch (bottom 
curve). Keeping baryon number fixed, the collapse of an
ordinary neutron star to its twin in the third family branch might release
huge amount of energy. A part of it may be carried away by gravitational waves.
This could be a good signal to probe interior of superdense stars. 
Damping times of superdense stars corresponding to the third family branch
have higher values than those of the neutron star branch.

We compare real and imaginary parts of eigenfrequencies of first axial w-modes
for EoS including $n$$p$$\Lambda$ and EoS involving $n$$p$$K^-$ with 
$U_{\bar K}(n_0) =$ -160 MeV in Figures 10 and 11. We find that values of real
and imaginary frequencies corresponding to both EoS are close beyond 
compactness 0.18. We conclude from these results and those recorded in Table I 
and II that these equations of state are indistinguishable in axial w-modes. 

\section{summary and conclusions}
We have studied quasi-normal modes of oscillating neutron stars with different
compositions and equations of state. This problem has been investigated 
adopting continued fraction method. We have computed frequencies and damping 
time scales of axial w-modes corresponding to EoS involving $n$$p$, 
$n$$p$$\Lambda$ and 
$n$$p$$K^-$ and $n$$p$H$K^-$$\bar K^0$ matter. We find that hyperons and/or 
antikaon condensates result softer EoS which ,in turn, lead to higher 
frequencies and lower damping time compared with those of $n$$p$ case. Axial
w-modes might distinguish between exotic matter and nucleons-only matter.
However, it would be difficult to differentiate the EoS of hyperon matter from
that of antikaon condensed matter using axial w-modes as probe. Further our
study has revealed that freuqencies of superdense stars in third family branch
are higher than their twins in the neutron star branch.

\newpage

\begin{table}
\begin{center}

\caption {First three values of characterisitic $l$=2 axial $w$-mode 
frequencies and damping timescales for neutron stars made of $n$, $p$,
$\Lambda$ and electrons and muons.
\label{tab1}}
\vskip 0.5 cm
\begin{tabular}{|ccccc|}
\hline
&&&&\\ 
{$M$}&{$R$}&{$\frac{M}{R}$}&{$\nu$}&{$\tau$} \\ 
&&&&\\ 
{($M_{\odot}$)}&{($km$)}&{}&{($kHz$)}&{($\mu s$)}\\ \hline
&&&&\\
{1.822}&{12.892}&{0.2}&{6.830}&{79.571}\\

{}&{}&{}&{10.055}&{47.711}\\
{}&{}&{}&{13.391}&{50.397}\\ \hline
&&&&\\

{1.867}&{12.325}&{0.22}&{6.796}&{87.129}\\
{}&{}&{}&{9.8556}&{52.440}\\
{}&{}&{}&{12.800}&{51.996}\\ \hline

&&&&\\
{1.885}&{11.514}&{0.24}&{6.800}&{94.766}\\
{}&{}&{}&{9.595}&{60.417}\\
{}&{}&{}&{12.148}&{54.029}\\ \hline

\end{tabular}
\end{center}
\end{table}
\newpage

\begin{table}
\begin{center}

\caption {The first three values of characterisitic $l$=2 axial $w$-mode 
frequencies and damping timescales for neutron stars undergoing a first order
phase transition from nuclear to $K^-$ condensed matter for antikaon optical 
potential depth at normal nuclear matter density $U_{\bar K}(n_0) = - 160 MeV$.
\label{tab2}}
\vskip 0.5 cm
\begin{tabular}{|ccccc|}
\hline
&&&&\\ 
{$M$}&{$R$}&{$\frac{M}{R}$}&{$\nu$}&{$\tau$} \\ 
&&&&\\ 
{($M_{\odot}$)}&{($km$)}&{}&{($kHz$)}&{($\mu s$)}\\ \hline
&&&&\\
{1.680}&{12.401}&{0.2}&{6.996}&{69.554}\\

{}&{}&{}&{10.107}&{46.401}\\
{}&{}&{}&{13.691}&{51.040}\\ \hline
&&&&\\

{1.770}&{11.641}&{0.22}&{6.941}&{80.302}\\
{}&{}&{}&{9.832}&{52.712}\\
{}&{}&{}&{12.779}&{53.119}\\ \hline

&&&&\\
{1.825}&{11.068}&{0.24}&{6.900}&{89.476}\\
{}&{}&{}&{9.593}&{60.270}\\
{}&{}&{}&{12.149}&{54.333}\\ \hline

\end{tabular}
\end{center}
\end{table}

\newpage
\vspace{-2cm}

{\centerline{
\epsfxsize=12cm
\epsfysize=14cm
\epsffile{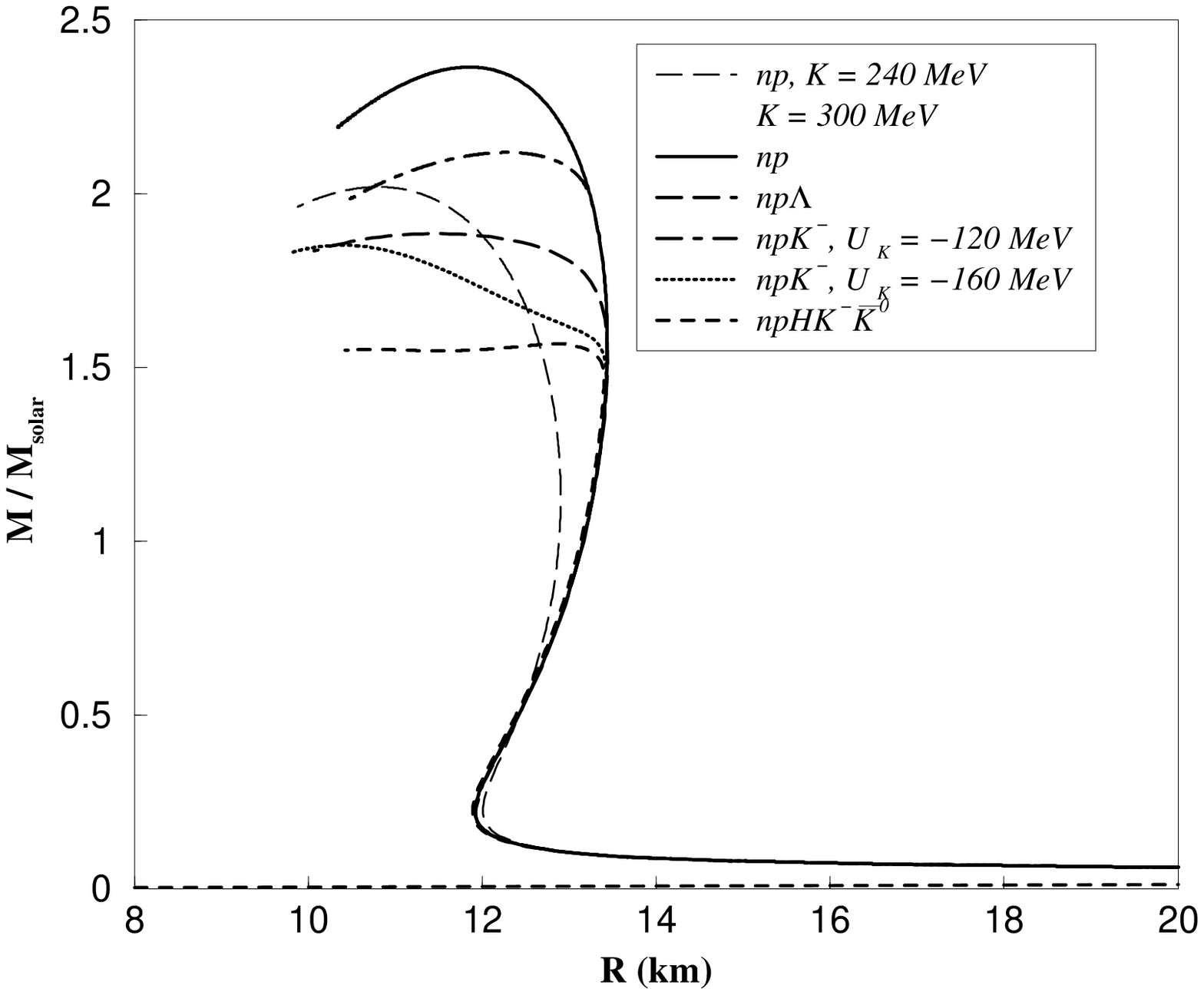}
}}
\noindent{\small{
Fig. 1. Gravitational mass is plotted with radius for compact stars having 
different compositions as explained in text.}}

\newpage
\vspace{-2cm}

{\centerline{
\epsfxsize=12cm
\epsfysize=14cm
\epsffile{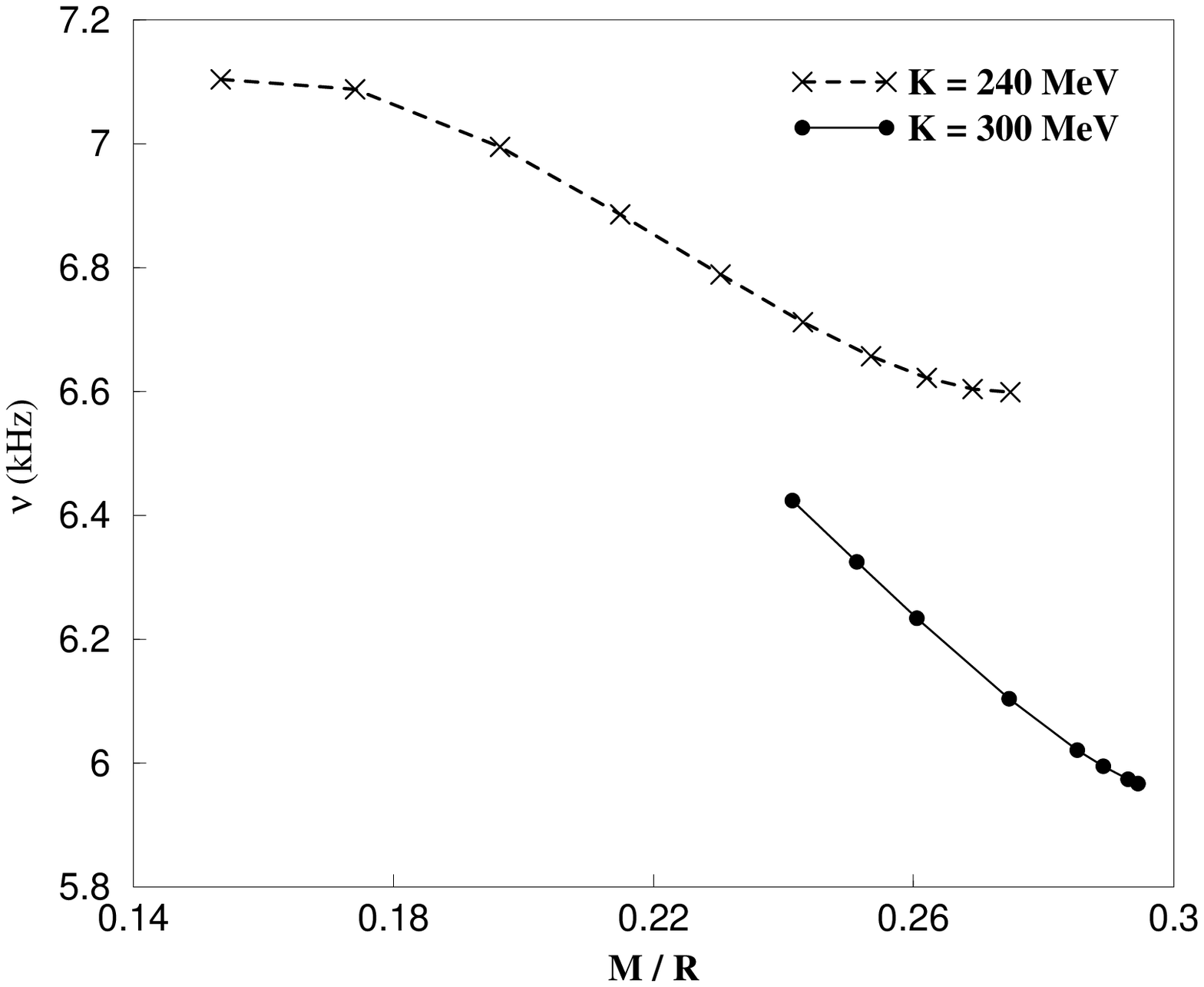}
}}
\noindent{\small{
Fig. 2. Frequency of first axial w-mode is plotted as a function of neutron 
star compactness for nucleons only matter corresponding to 
incompressibility K=240 and 300 MeV.}}

\newpage
\vspace{-2cm}

{\centerline{
\epsfxsize=12cm
\epsfysize=14cm
\epsffile{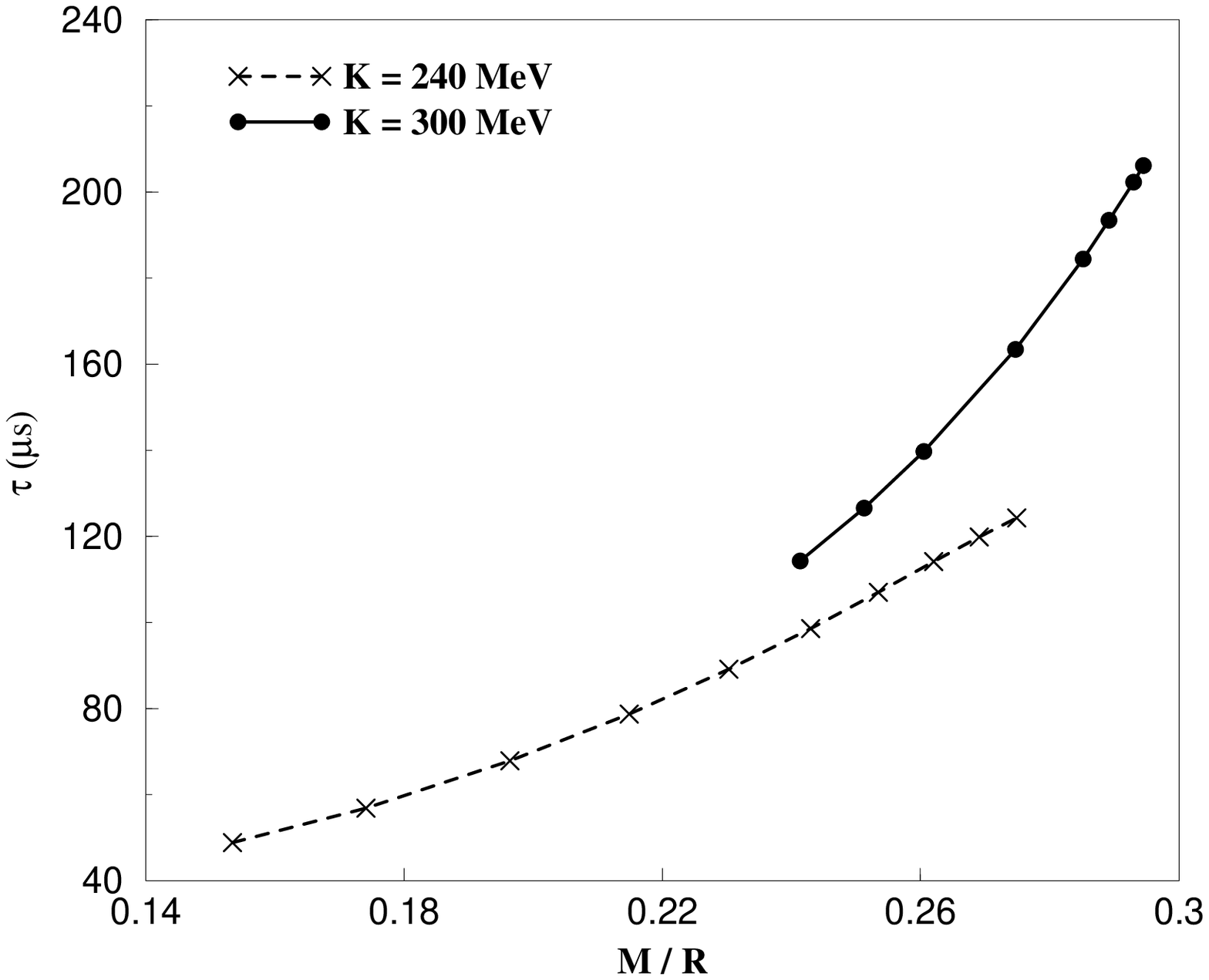}
}}
\noindent{\small{
Fig. 3. Damping time of first axial w-mode is plotted as a function of 
neutron star compactness for nucleons only matter corresponding to 
incompressibility K=240 and 300 MeV.}}
\newpage
\vspace{-2cm}

{\centerline{
\epsfxsize=12cm
\epsfysize=14cm
\epsffile{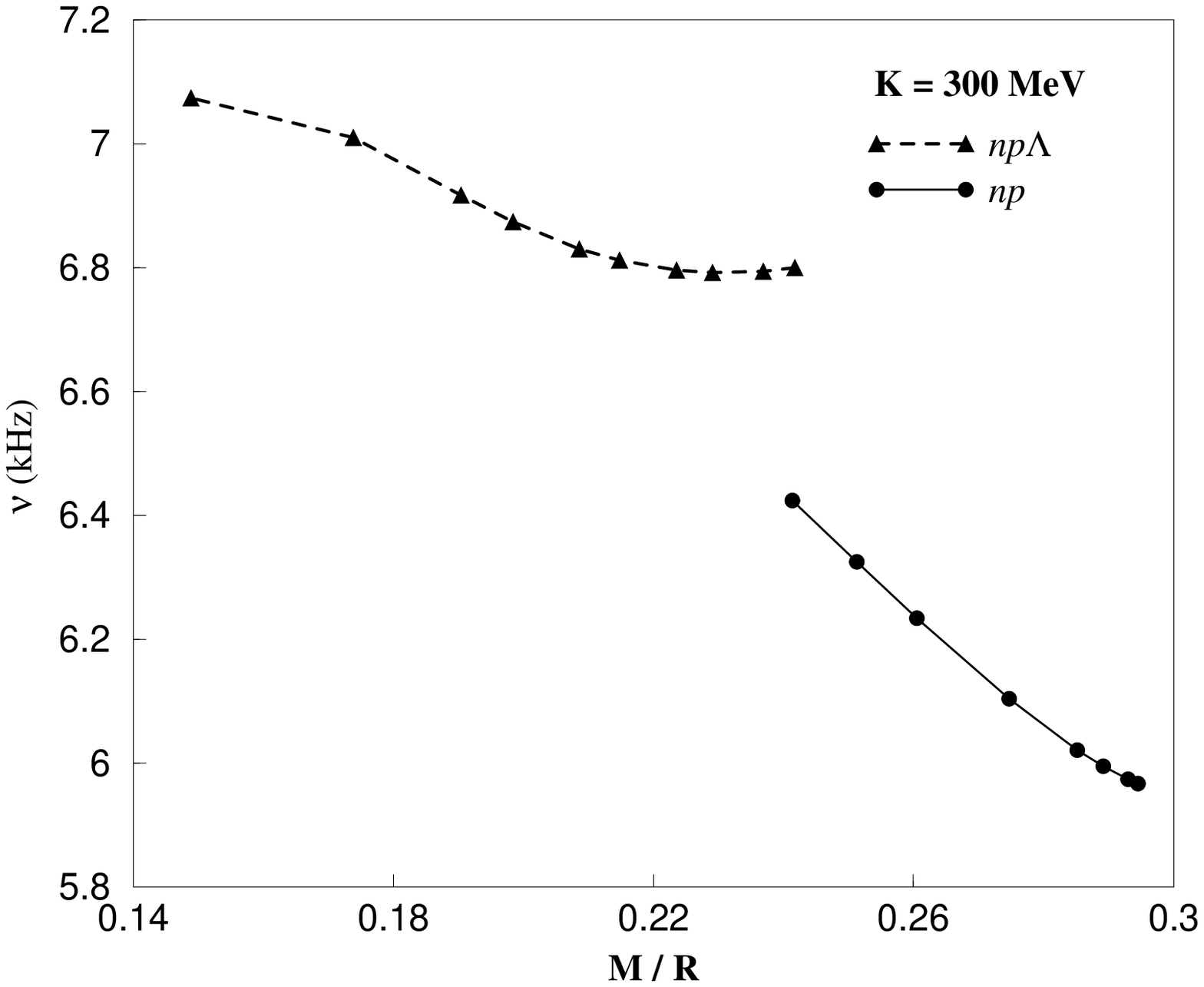}
}}

\noindent{\small{
Fig. 4. Frequency of first axial w-mode is plotted as a function of neutron 
star compactness for $n$$p$ as well as $n$$p$$\Lambda$ matter 
with K=300 MeV.}}
\newpage
\vspace{-2cm}

{\centerline{
\epsfxsize=12cm
\epsfysize=14cm
\epsffile{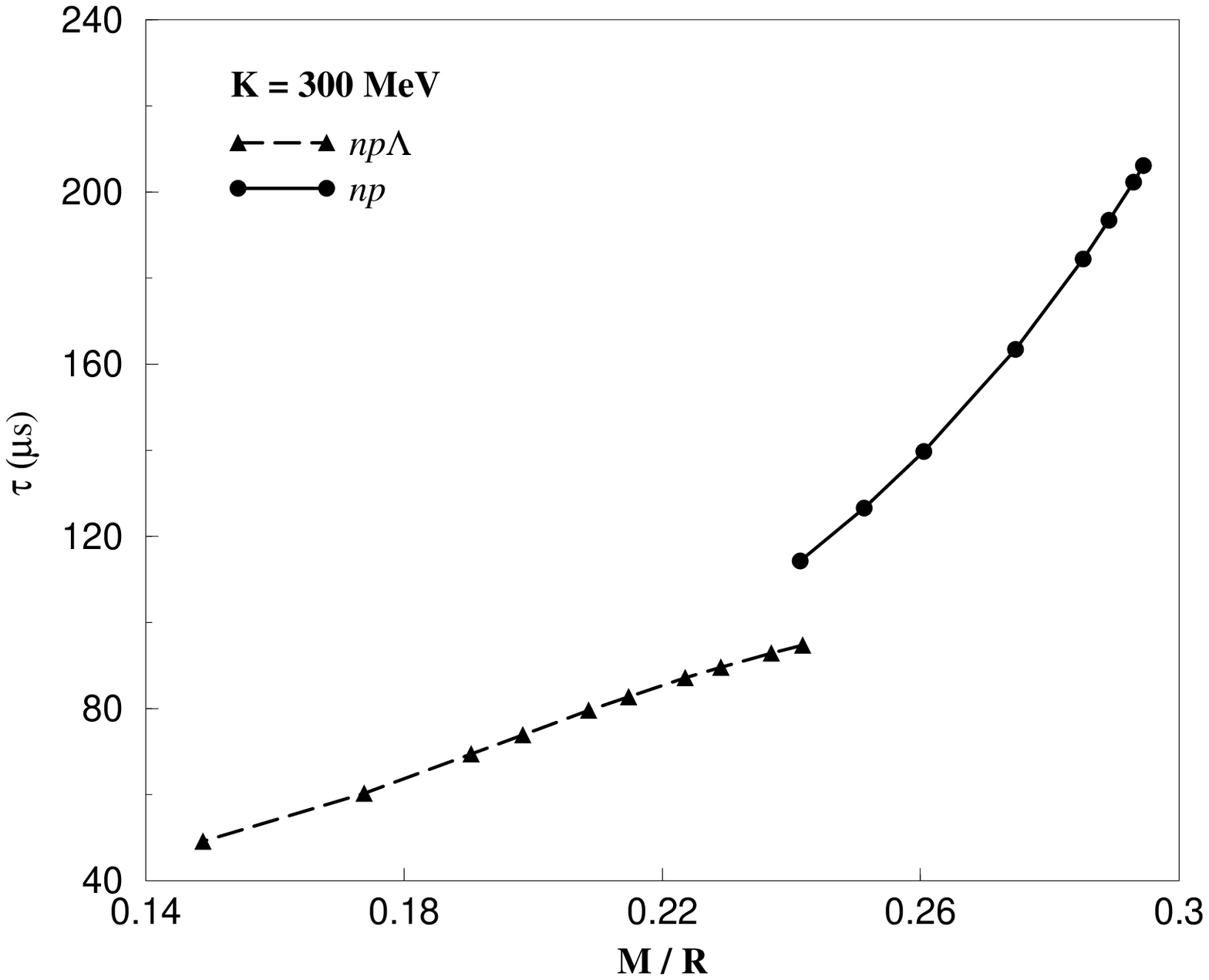}
}}

\noindent{\small{
Fig. 5. Damping time of first axial w-mode is plotted as a function of 
neutron star compactness for $n$$p$ as well as $n$$p$$\Lambda$ 
matter with K=300 MeV.}}

\newpage
\vspace{-2cm}

{\centerline{
\epsfxsize=12cm
\epsfysize=14cm
\epsffile{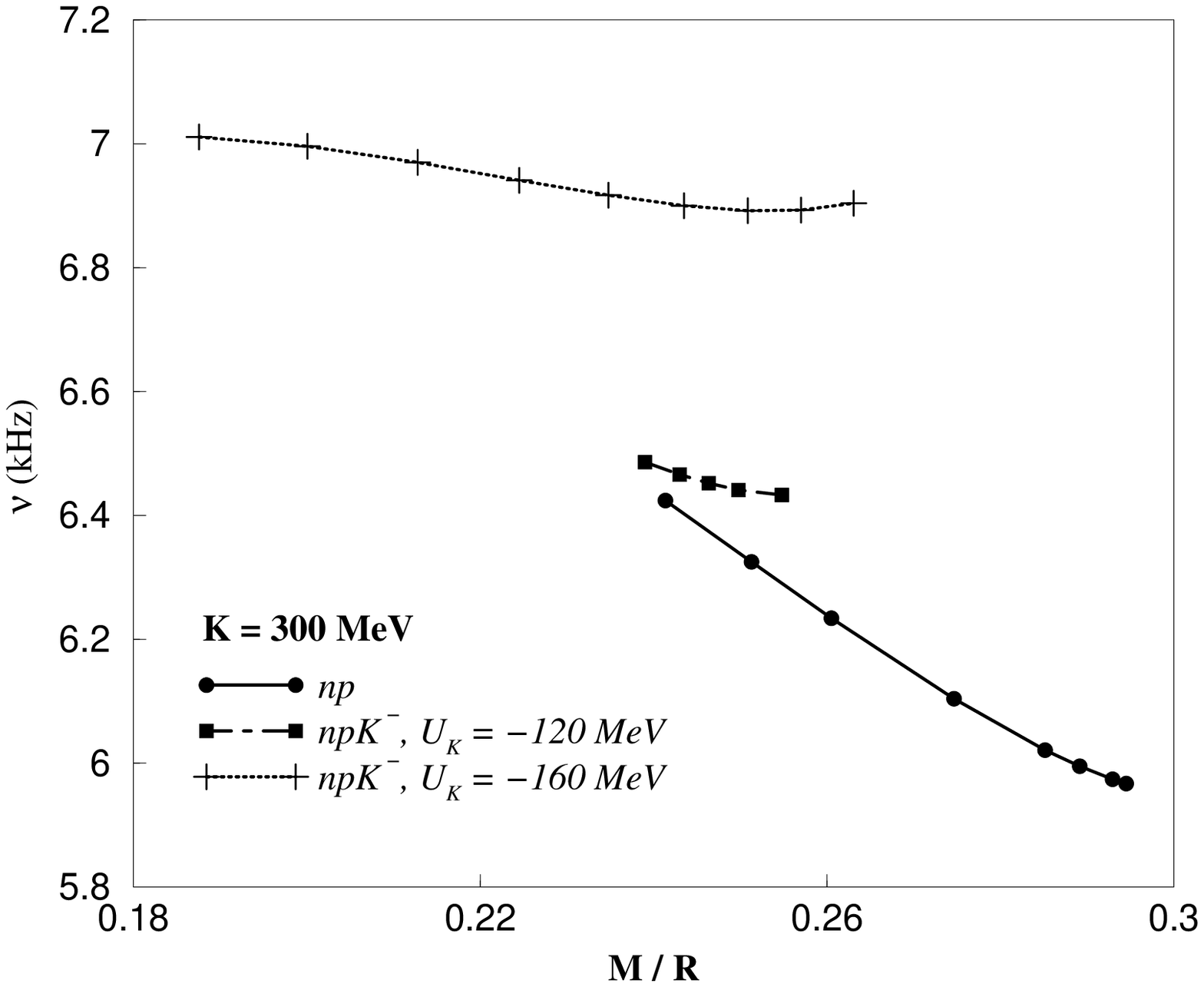}
}}

\noindent{\small{
Fig. 6. Frequency of first axial w-mode is plotted as a function of 
neutron star compactness for $n$$p$ and $n$$p$$K^-$ 
matter with K=300 MeV and $U_{\bar K}(n_0)$ = -120, -160 MeV.}}

\newpage
\vspace{-2cm}

{\centerline{
\epsfxsize=12cm
\epsfysize=14cm
\epsffile{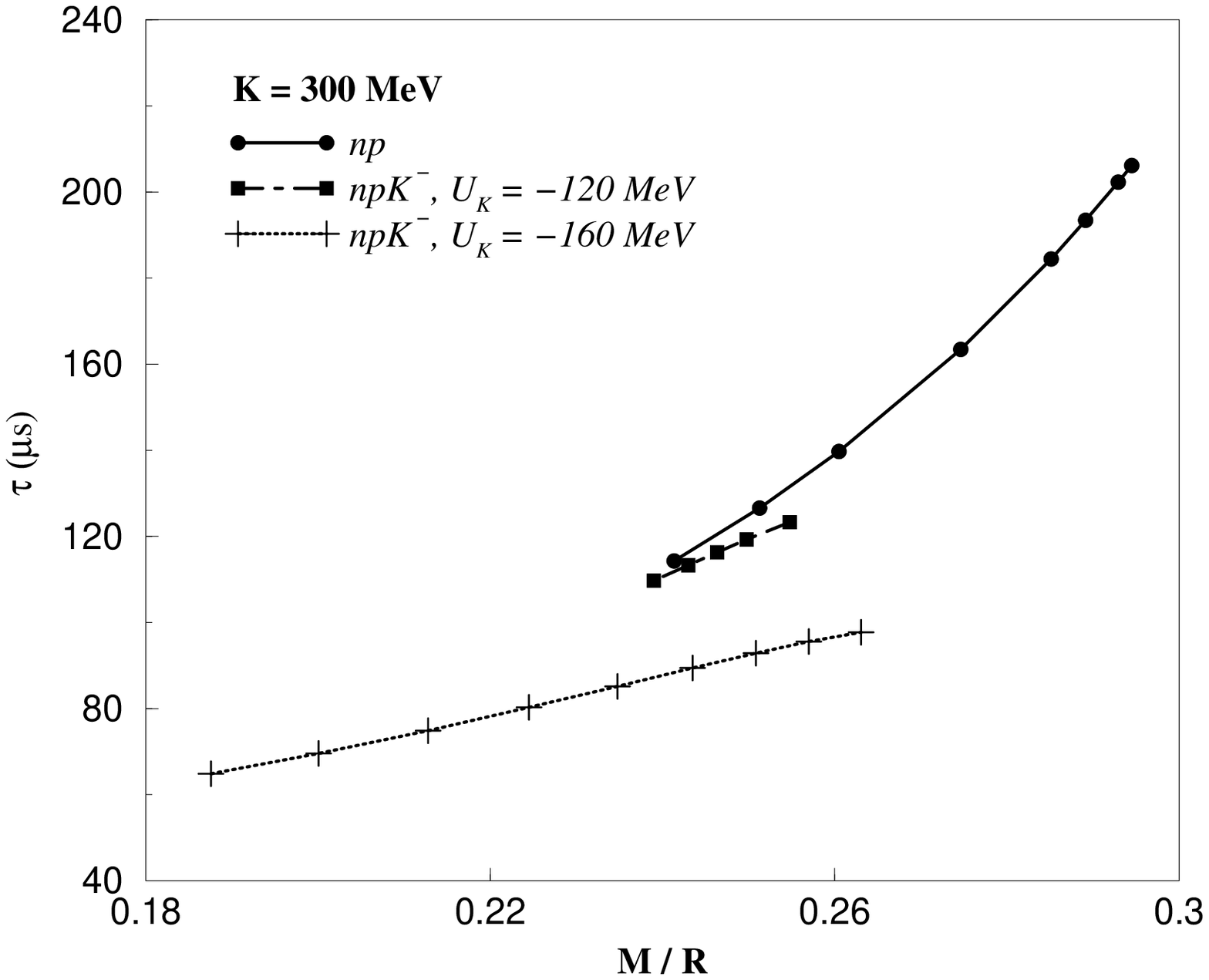}
}}

\noindent{\small{
Fig. 7. Damping time of first axial w-mode is plotted as a function of 
neutron star compactness for $n$$p$ and $n$$p$$K^-$ 
matter with K=300 MeV and $U_{\bar K}(n_0)$ = -120, -160 MeV.}}
\newpage
\vspace{-2cm}

{\centerline{
\epsfxsize=12cm
\epsfysize=14cm
\epsffile{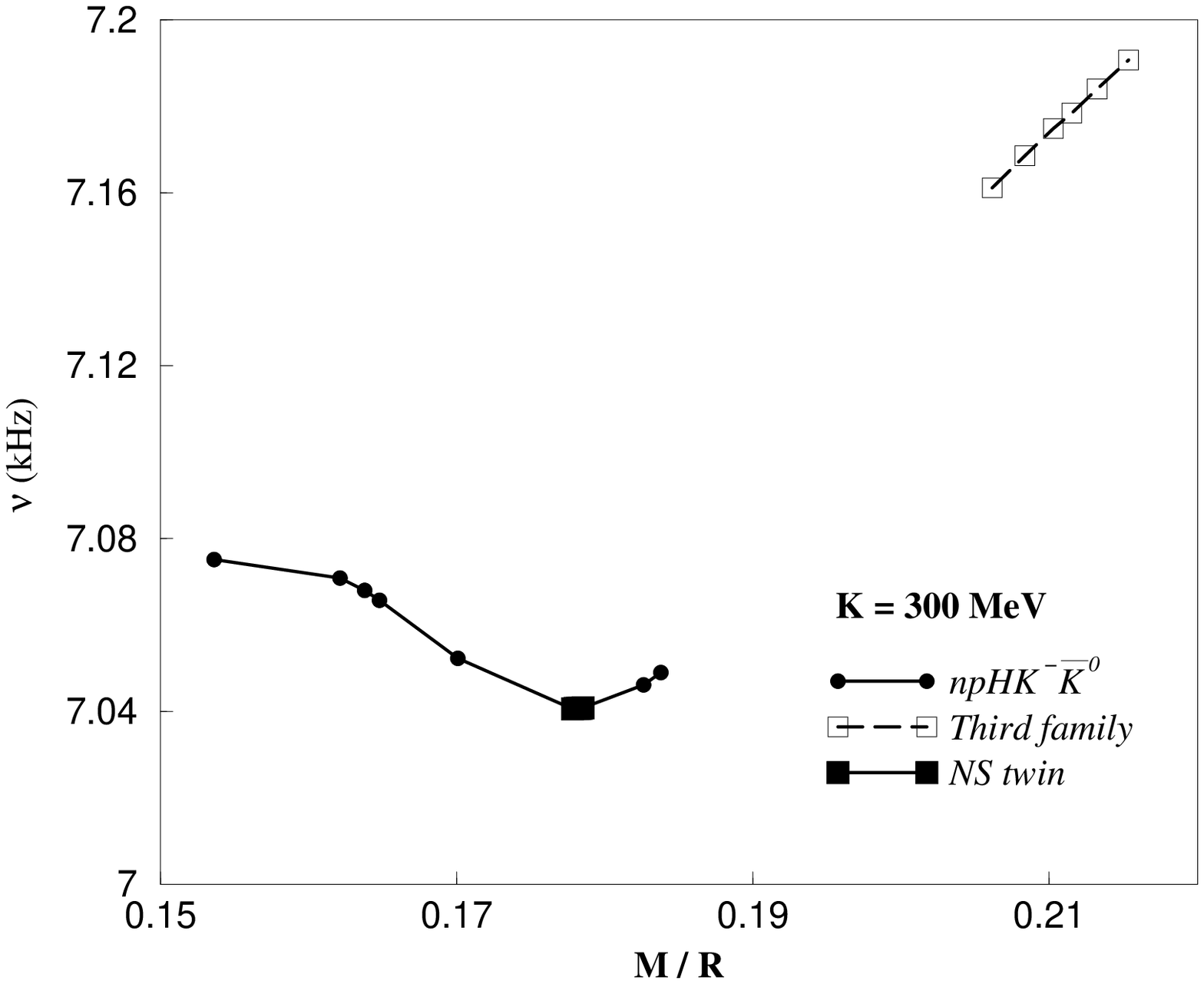}
}}

\noindent{\small{
Fig. 8. Frequency of first axial w-mode is plotted as a function of 
compactness for neutron stars including $n$,$p$,$\Lambda$,$\Xi^-$,$\Xi^0$,
$\Sigma^-$,$K^-$ and $\bar K^0$ with K=300 MeV and 
$U_{\bar K}(n_0)$ = -160 MeV. The lower curve shows the neutron star branch 
whereas the third family branch is shown by the upper curve. Neutron star twins 
corresponding to the superdense stars in the third family branch are denoted by
filled squares in the lower curve.}}
\newpage
\vspace{-2cm}

{\centerline{
\epsfxsize=12cm
\epsfysize=14cm
\epsffile{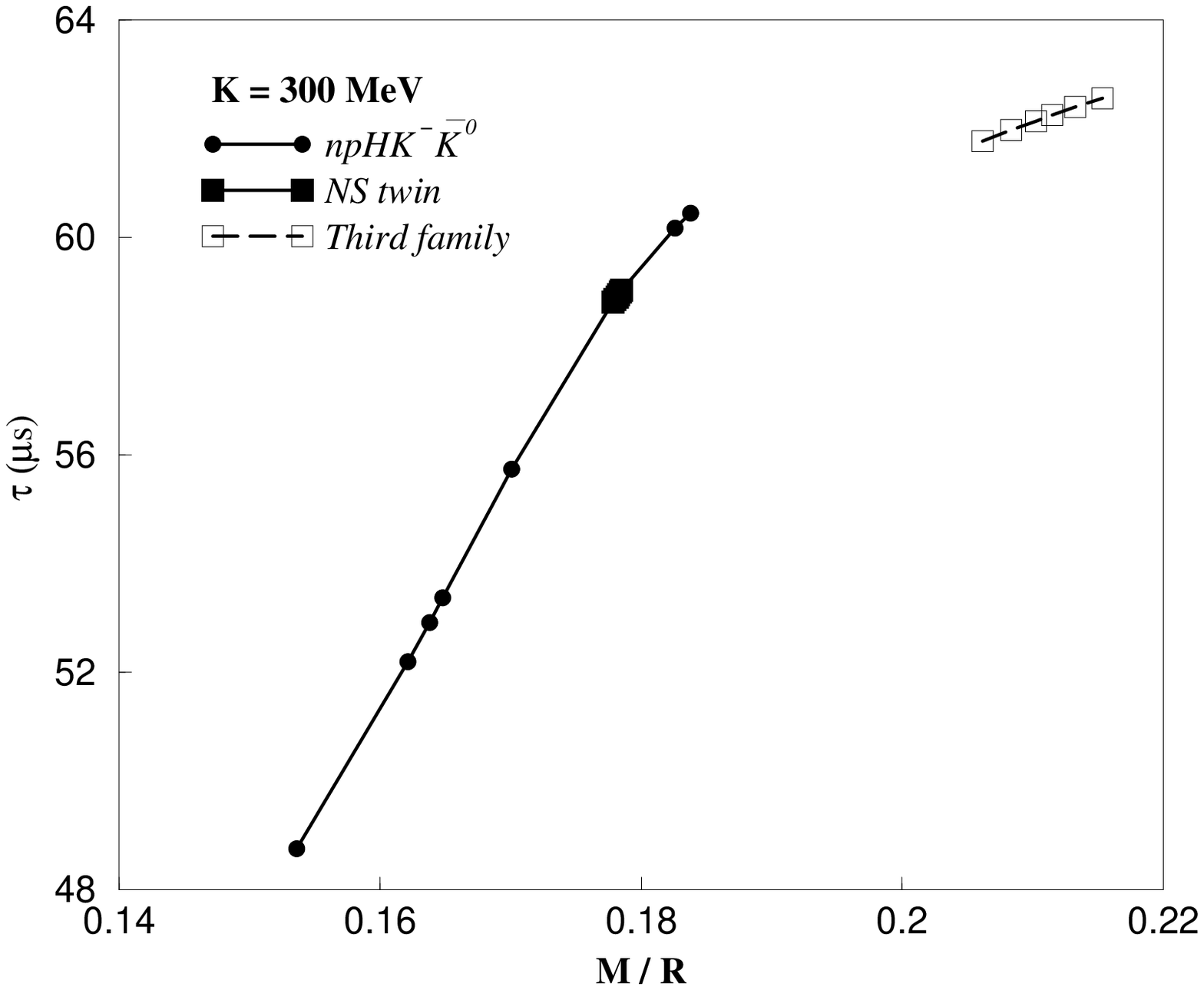}
}}

\noindent{\small{
Fig. 9. Damping time of first axial w-mode is plotted as a function of 
compactness for neutron stars including $n$,$p$,$\Lambda$,$\Xi^-$,$\Xi^0$,
$\Sigma^-$,$K^-$ and $\bar K^0$ with K=300 MeV and 
$U_{\bar K}(n_0)$ = -160 MeV. The lower curve shows the neutron star branch 
whereas the third family branch is shown by the upper curve. Neutron star twins 
corresponding to the superdense stars in the third family branch are denoted by
filled squares in the lower curve.}}
\newpage
\vspace{-2cm}

{\centerline{
\epsfxsize=12cm
\epsfysize=14cm
\epsffile{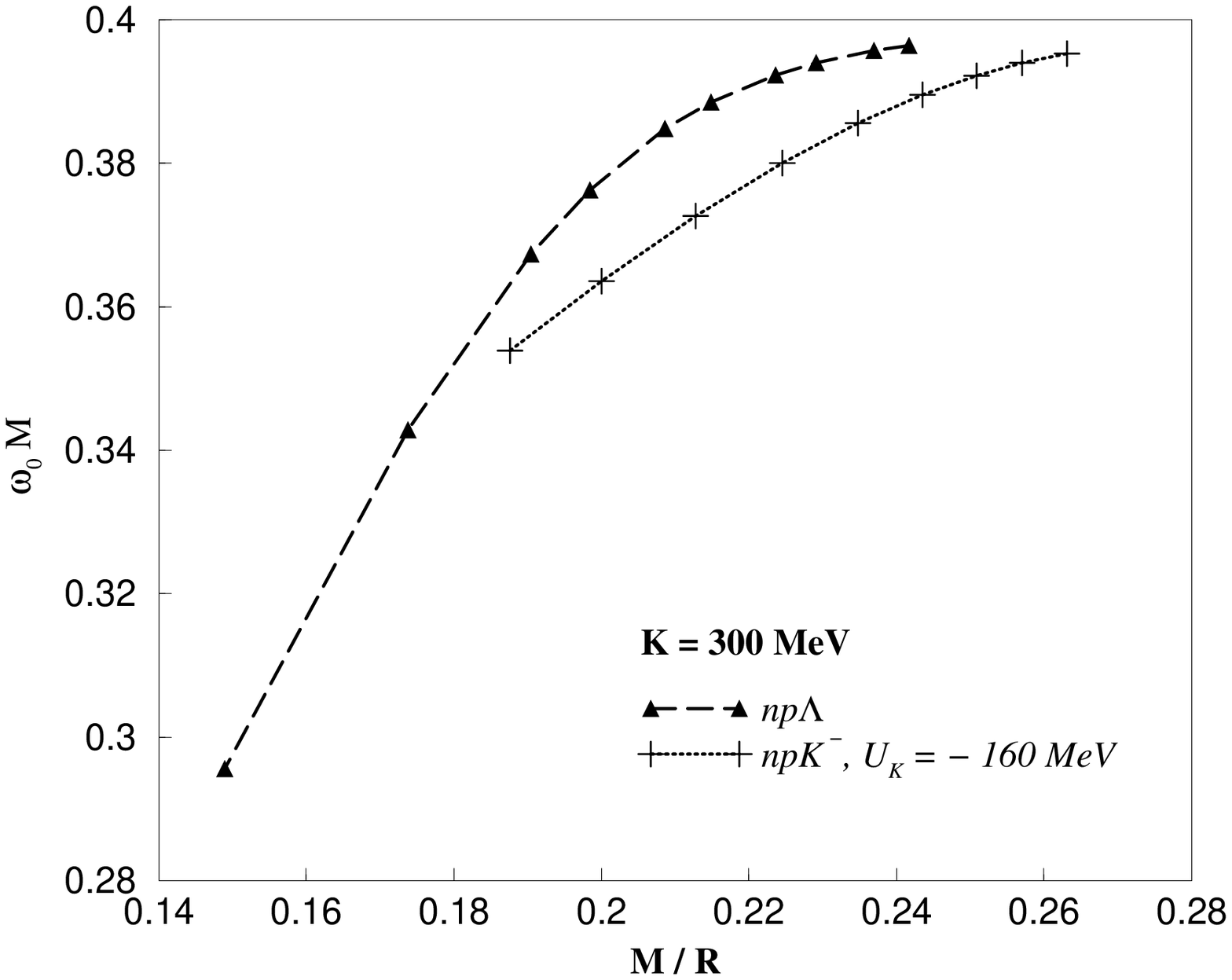}
}}

\noindent{\small{
Fig. 10. Real part of eigenfrequency of first axial w-mode is plotted as 
a function of compactness for $n$$p$$\Lambda$ and 
$n$$p$$K^-$ matter with K=300 MeV and 
$U_{\bar K}(n_0)$ = -160 MeV.}}
\newpage
\vspace{-2cm}

{\centerline{
\epsfxsize=12cm
\epsfysize=14cm
\epsffile{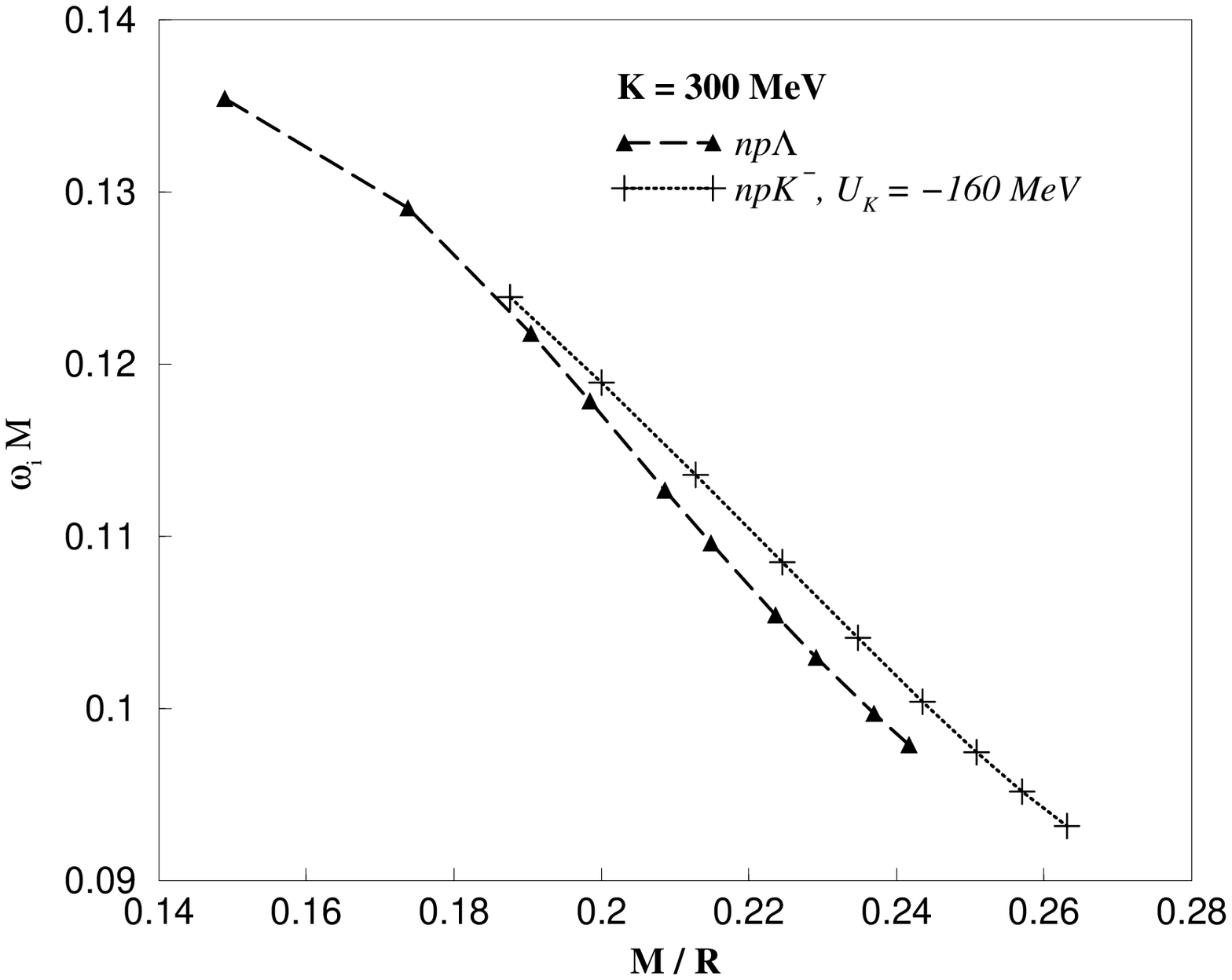}
}}

\noindent{\small{
Fig. 11. Imaginary part of eigenfrequency of first axial w-mode is plotted as 
a function of compactness for $n$$p$$\Lambda$ and 
$n$$p$$K^-$ matter with K=300 MeV and $U_{\bar K}(n_0)$ = -160 MeV.}}
\end{document}